\documentclass[twocolumn,pre,superscriptaddress]{revtex4}

\usepackage{dcolumn}
\usepackage{amsmath}

\usepackage{graphicx}

\begin{document}

\renewcommand{\d}{\mathrm{d}}
\newcommand{\Ord}{\mathrm{O}}
\newcommand{\e}{\mathrm{e}}
\newcommand{\half}{\mbox{$\frac12$}}
\newcommand{\set}[1]{\lbrace#1\rbrace}
\newcommand{\av}[1]{\langle#1\rangle}
\newcommand{\etal}{{\it{}et~al.}}
\newcommand{\defn}{\textit}
\newcommand{\Beta}{\mathrm{B}}

\newlength{\figurewidth}
\setlength{\figurewidth}{0.95\columnwidth}
\setlength{\parskip}{0pt}
\setlength{\tabcolsep}{6pt}
\setlength{\arraycolsep}{2pt}

\title{Exact solutions for models of evolving networks with addition and
deletion of nodes}
\author{Cristopher Moore}
\affiliation{Department of Computer Science and Department of Physics
and Astronomy,\\
University of New Mexico, Albuquerque, NM 87131}
\affiliation{Santa Fe Institute, Santa Fe NM 87501}
\affiliation{Center for the Study of Complex Systems, University of
Michigan, Ann Arbor, MI 48109}
\author{Gourab Ghoshal}
\affiliation{Department of Physics, University of Michigan, Ann Arbor, MI
48109}
\affiliation{Michigan Center for Theoretical Physics, University of
Michigan, Ann Arbor, MI, 48109}
\author{M. E. J. Newman}
\affiliation{Center for the Study of Complex Systems, University of
Michigan, Ann Arbor, MI 48109}
\affiliation{Department of Physics, University of Michigan, Ann Arbor, MI
48109}

\begin{abstract}
There has been considerable recent interest in the properties of networks,
such as citation networks and the worldwide web, that grow by the addition
of vertices, and a number of simple solvable models of network growth have
been studied.  In the real world, however, many networks, including the
web, not only add vertices but also lose them.  Here we formulate models of
the time evolution of such networks and give exact solutions for a number
of cases of particular interest.  For the case of net growth and so-called
preferential attachment---in which newly appearing vertices attach to
previously existing ones in proportion to vertex degree---we show that the
resulting networks have power-law degree distributions, but with an
exponent that diverges as the growth rate vanishes.  We conjecture that the
low exponent values observed in real-world networks are thus the result of
vigorous growth in which the rate of addition of vertices far exceeds the
rate of removal.  Were growth to slow in the future, for instance in a more
mature future version of the web, we would expect to see exponents
increase, potentially without bound.
\end{abstract}
\pacs{89.75.Hc, 87.23.Ge, 89.20.Hh, 05.10.-a}
\maketitle

\section{Introduction}
The study of networks has attracted a substantial amount of attention from
the physics community in the last few years~\cite{AB02,DM02,Newman03d}, in
part because of networks' broad utility as representations of real-world
complex systems and in part because of the demonstrable successes of
physics techniques in shedding light on networked phenomena.  One topic
that has been the subject of a particularly large volume of work is growing
networks, such as citation networks~\cite{Price65,Redner98} and the
worldwide web~\cite{AJB99,Kleinberg99b}.  Perhaps the best-known body of
work on this topic is that dealing with ``preferential attachment''
models~\cite{Price76,BA99b}, in which vertices are added to a network with
edges that attach to preexisting vertices with probabilities depending on
those vertices' degrees.  When the attachment probability is precisely
linear in the degree of the target vertex the resulting degree sequence for
the network follows a Yule distribution in the limit of large network size,
meaning it has a power-law tail~\cite{Price76,BA99b,KRL00,DMS00,BRST01}.
This case is of special interest because both citation networks and the
worldwide web are observed to have degree distributions that approximately
follow power laws.

The preferential attachment model may be quite a good model for citation
networks, which is one of the cases for which it was originally
proposed~\cite{Price76,KRL00}.  For other networks, however, and especially
for the worldwide web, it is, as many authors have pointed out, necessarily
incomplete~\cite{DM00b,AB00a,KR02a,Tadic02,GSM05}.  On the web there are
clearly other processes taking place in addition to the deposition of
vertices and edges.  In particular, it is a matter of common experience
that vertices (i.e.,~web pages) are often removed from the web, and with
them the links that they had to other pages.  Models of this process have
been touched upon occasionally in the literature~\cite{CL04,CFV04} and the
evidence suggests that in some cases vertex deletion affects the crucial
power-law behavior of the degree distribution, while in other cases it does
not.

In this paper, we study the general process in which a network grows (or,
potentially, shrinks) by the constant addition and removal of vertices and
edges.  We show that a class of such processes can be solved exactly for
the degree distributions they generate by solving differential equations
governing the probability generating functions for those distributions.  In
particular, we give solutions for three example problems of this type,
having uniform or preferential attachment, and having stationary size or
net growth.  The case of uniform attachment and stationary size is of
interest as a possible model for the structure of peer-to-peer filesharing
networks, while the preferential-attachment stationary-size case displays a
nontrivial stretched exponential form in the tail of the degree
distribution.  Our solution of the preferential attachment case with net
growth confirms earlier results indicating that this process generates a
power-law distribution, although the exponent of the power-law diverges as
the growth rate tends to zero, giving degree distributions that are
numerically indistinguishable from exponential for small growth rates.
This suggests that the clear power law seen in the real worldwide web is a
signature of a network whose rate of vertex accrual far outstrips the rate
at which vertices are removed.  The relative rates of addition and removal
could, however, change as the web matures, possibly leading to a loss of
power-law behavior at some point in the future.

\section{The model}
Consider a network that evolves by the addition and removal of vertices.
In each unit of time, we add a single vertex to the network and remove~$r$
vertices.  When a vertex is removed so too are all the edges incident on
that vertex, which means that the degrees of the vertices at the other ends
of those edges will decrease.  Non-integer values of~$r$ are permitted and
are interpreted in the usual stochastic fashion.  (For example, values
$r<1$ can be interpreted as the probability per unit time that a vertex is
removed.)  The value $r=1$ corresponds to a network of fixed size in which
there is vertex turnover but no growth; values $r<1$ correspond to growing
networks.  In principle one could also look at values $r>1$, which
correspond to shrinking networks, and the methods derived here are
applicable to that case.  However, we are not aware of any real-world
examples of shrinking networks in which the asymptotic degree distribution
is of interest, so we will not pursue the shrinking case here.

We make two further assumptions, which have also been made by most previous
authors in studying these types of systems: (1)~that all vertices added
have the same initial degree, which we denote~$c$; (2)~that the vertices
removed are selected uniformly at random from the set of all extant
vertices.  Note however that we will not assume that the network is
uncorrelated (i.e.,~that it is a random multigraph conditioned on its
degree distribution as in the so-called configuration model).  In general
the networks we consider will have correlations among the degrees of their
vertices but our solutions will nonetheless be exact.

Let $p_k$ be the fraction of vertices in the network at a given time that
have degree~$k$.  By definition, $p_k$~has the normalization
\begin{equation}
\label{eq:normpk}
\sum_{k=0}^\infty p_k = 1.
\end{equation}
Our primary goal in this paper will to evaluate exactly the degree
distribution~$p_k$ for various cases of interest.

Although the form of $p_k$ is, as we will see, highly nontrivial in most
cases, the mean degree of a vertex $\av{k}=\sum_{k=0}^\infty kp_k$ is
easily derived in terms of the parameters~$r$ and~$c$.  The mean number of
vertices added to the network per unit time is $1-r$.  The mean number of
edges removed when a randomly chosen vertex is removed from the network is
by definition~$\av{k}$.  Thus the mean number of edges added to the network
per unit time is $c-r\av{k}$.  For a graph of $m$ edges and $n$ vertices,
the mean degree is $\av{k}=2m/n$.  After time~$t$ we have $n=(1-r)t$ and,
assuming that $\av{k}$ has an asymptotically constant value,
$m=(c-r\av{k})t$.  Thus
\begin{equation}
\av{k} = 2 {c-r\av{k}\over1-r},
\end{equation}
or, rearranging,
\begin{equation}
\av{k} = {2c\over1+r}.
\label{eq:avk}
\end{equation}
In the special case $r=1$ of a constant-size network, this gives
$\av{k}=c$, which is clearly the correct answer.

We must also consider how an added vertex chooses the $c$ other vertices to
which it attaches.  Let us define the attachment kernel~$\pi_k$ to be $n$
times the probability that a given edge of a newly added vertex attaches to
a given preexisting vertex of degree~$k$.  The factor of $n$ here is
convenient, since it means that the total probability that the given edge
attaches to any vertex of degree~$k$ is simply $\pi_kp_k$.  Since each edge
must attach to a vertex of \emph{some} degree, this also immediately
implies that the correct normalization for $\pi_k$ is
\begin{equation}
\label{eq:normpik}
\sum_{k=0}^\infty \pi_k p_k = 1.
\end{equation}

For the particular case of $\pi_k\propto k$ and $r<1$, which we consider in
Section~\ref{sec:prefgrow}, models similar to ours have been studied
previously by Chung and Lu~\cite{CL04} and by Cooper, Frieze, and
Vera~\cite{CFV04}.  The results reported by these authors are mostly of a
different nature to ours, but there are some overlaps, which we discuss at
the appropriate point.

\subsection{Rate equation}
\label{sec:rateeq}
Given these definitions, the evolution of the degree distribution is
governed by a rate equation as follows.  If there are at total of $n$
vertices in the network at a given time then the number of vertices with
degree~$k$ is~$np_k$.  One unit of time later this number is $(n+1-r)p_k'$,
where $p_k'$ is the new value of~$p_k$.  Then
\begin{eqnarray}
\label{eq:dpk}
(n+1-r)p_k' &=& np_k + \delta_{kc} + c\pi_{k-1} p_{k-1} - c\pi_k p_k
                \nonumber\\
            & & {} + r (k+1) p_{k+1} - r k p_k - r p_k .
\end{eqnarray}
The term $\delta_{kc}$ in Eq.~\eqref{eq:dpk} represents the addition of a
vertex of degree~$c$ to the network.  The terms $c \pi_{k-1} p_{k-1}$ and
$-c \pi_k p_k$ describe the flow of vertices from degree $k-1$ to~$k$ and
from $k$ to $k+1$ as they gain extra edges when newly added vertices attach
to them.  The terms $(k+1) p_{k+1}$ and $-k p_k$ describe the flow from
degree $k+1$ to~$k$ and from $k$ to $k-1$ as vertices loose edges when one
of their neighbors is removed from the network.  And the term $-rp_k$
represents the removal with probability~$r$ of a vertex with degree~$k$.
Contributions from processes in which a vertex gains or loses two or more
edges in a single unit of time vanish in the limit of large~$n$ and have
been neglected.

We will be interested in the asymptotic form of $p_k$ in the limit of large
times for a given~$\pi_k$.  Setting $p_k'=p_k$ in~\eqref{eq:dpk} gives
\begin{eqnarray}
\label{eq:recur}
\delta_{kc} &+& c\pi_{k-1} p_{k-1} - c\pi_k p_k \nonumber\\
            &+& r (k+1) p_{k+1} - r k p_k - p_k = 0,
\end{eqnarray}
whose solution we can write in terms of generating functions as follows.
Let us define
\begin{eqnarray}
f(z) &=& \sum_{k=0}^\infty \pi_k p_k z^k, \\
g(z) &=& \sum_{k=0}^\infty p_k z^k.
\end{eqnarray}
Then, upon multiplying both sides of~\eqref{eq:recur} by $z^k$ and summing
over~$k$ (with the convention that $p_{-1}=0$), we derive a differential
equation for $g(z)$ thus:
\begin{equation}
r(1-z) {\d g\over\d z} - g(z) - c(1-z) f(z) + z^c = 0.
\label{eq:fg}
\end{equation}
Note also that we can easily generalize our model to the case where the
degrees of the vertices added are not all identical but are instead drawn
at random from some distribution~$r_k$.  In that case, we simply replace
$\delta_{kc}$ in Eq.~\eqref{eq:recur} with $r_k$ and $z^c$ in
Eq.~\eqref{eq:fg} with the generating function $h(z) = \sum_k r_k z^k$.

In the following sections we solve Eq.~\eqref{eq:fg} for a number of
different choices of the attachment kernel~$\pi_k$.  Note that, since the
definitions of both $f(z)$ and $g(z)$ incorporate the unknown
distribution~$p_k$, we must in general solve implicitly for $g(z)$ in terms
of~$f(z)$.  In all of the cases of interest to us here, however, it turns
out to be straightforward to derive a explicit equation for $g(z)$ as a
special case of~\eqref{eq:fg}.

\section{Solutions for specific cases}
In this section we study three specific examples of the class of models
defined in the preceding section, namely linear preferential attachment
models ($\pi_k\propto k$) for both growing and fixed-size networks, and
uniform attachment ($\pi_k=\mbox{constant}$) for fixed size.  As we will
see, each of these cases turns out to have interesting features.

\subsection{Uniform attachment and constant size}
For the first of our example models we study the case where the size of the
network is constant ($r=1$) and in which each vertex added chooses the $c$
others to which it attaches uniformly at random.  This means that $\pi_k$
is constant, independent of~$k$, and, combining Eqs.~\eqref{eq:normpk}
and~\eqref{eq:normpik}, we immediately see that the correct normalization
for the attachment kernel is $\pi_k=1$ for all~$k$.  Then we have $\pi_k
p_k = p_k$ so that $f(z)=g(z)$ in~\eqref{eq:fg}, which gives
\begin{equation}
\biggl( c + {1\over1-z} \biggr) g(z) - {\d g\over\d z} = {z^c\over1-z}.
\label{eq:uniconst}
\end{equation}
Noting that $(1-z)\e^{-cz}$ is an integrating factor and that $g(z)$ must
obey the boundary condition $g(1)=1$, we readily determine that
\begin{eqnarray}
g(z) &=& {\e^{cz}\over1-z} \int_z^1 t^c \e^{-ct} \>\d t \nonumber\\
     &=& \frac{\e^{cz}}{1-z} c^{-(c+1)}
         \bigl[ \Gamma(c+1,cz) - \Gamma(c+1,c) \bigr],\quad
\label{eq:solngz1}
\end{eqnarray}
where 
\begin{equation}
\Gamma(c+1,x) = \int_x^\infty t^c \e^{-t} \>\d t.
\end{equation}
is the incomplete $\Gamma$-function.

One can easily check that this gives a mean degree $g'(1) = c$, as it must,
and that the variance of the degree $g''(1)+g'(1)-c^2$ is equal to
$\frac{2}{3}c$, indicating a tightly peaked degree distribution.

To obtain an explicit expression for the degree distribution, we make use
of
\begin{eqnarray}
\label{eq:expgamma}
\Gamma(c+1,x) &=& \Gamma(c+1) \,\e^{-x} \sum_{m=0}^c \frac{x^m}{m!}, \\
\e^x          &=& \sum_{m=0}^\infty \frac{x^m}{m!}, \\
(1-z)^{-1}    &=& \sum_{k=0}^\infty z^k,
\end{eqnarray}
to write
\begin{eqnarray}
g(z) &=& c^{-(c+1)} \nonumber\\
     & & \hspace{-3em}{}\times\sum_{k=0}^\infty z^k
         \biggl[ \Gamma(c+1) \sum_{m=0}^c {(cz)^m\over m!}
         - \Gamma(c+1,c) \sum_{m=0}^\infty {(cz)^m\over m!} \biggr].
         \nonumber\\
\label{eq:gzrewrite}
\end{eqnarray}
The $z$-dependence in the first term of this expression can be rewritten
\begin{eqnarray}
\sum_{k=0}^\infty z^k \sum_{m=0}^c {(cz)^m\over m!}
  &=& \sum_{m=0}^c \sum_{k=m}^\infty z^k {c^m\over m!} \nonumber\\
  &=& \sum_{k=0}^\infty z^k \sum_{m=0}^{\min(k,c)} {c^m\over m!}
      \nonumber\\
  &=& \e^c \sum_{k=0}^\infty z^k {\Gamma\bigl(\min(k,c)+1,c\bigr)\over
      \Gamma\bigl(\min(k,c)+1\bigr)},\quad{}
\end{eqnarray}
where $\min(k,c)$ denotes the smaller of $k$ and $c$ and we have again
employed~\eqref{eq:expgamma}.  A similar sequence of manipulations leads to
an expression for the second term also, thus:
\begin{equation}
\sum_{k=0}^\infty z^k \sum_{m=0}^\infty {(cz)^m\over m!}
  = \e^c \sum_{k=0}^\infty z^k {\Gamma(k+1,c)\over\Gamma(k+1)}.
\end{equation}
Combining these identities with~\eqref{eq:gzrewrite}, it is then a simple
matter to read off the term in $g(z)$ involving~$z^k$, which is by
definition our~$p_k$.  We find two separate expressions for the cases of
$k$ above or below~$c$:
\begin{eqnarray}
p_k &=& {\e^c\over c^{c+1}}
      \bigl[ \Gamma(c+1) - \Gamma(c+1,c) \bigr]
      {\Gamma(k+1,c)\over\Gamma(k+1)},\nonumber\\
    & & \hspace{13em}\mbox{for $k<c$},
\label{eq:pkkltc}
\end{eqnarray}
and
\begin{equation}
p_k = {\e^c\over c^{c+1}}
      \Gamma(c+1,c) \biggl[ 1 - 
      {\Gamma(k+1,c)\over\Gamma(k+1)} \biggr],\quad\mbox{for $k\ge c$}.
\label{eq:pkkgtc}
\end{equation}
Note that the quantity $\Gamma(k+1,c)/\Gamma(k+1)$ appearing in both these
expressions is the probability that a Poisson-distributed variable with
mean $c$ is less than or equal to~$k$.  Thus the degree distribution has a
tail that decays as the cumulative distribution of such a Poisson variable,
implying that it falls off rapidly.  To see this more explicitly, we note
that for fixed~$c$ and $k\gg c$
\begin{equation}
p_k = {\Gamma(c+1,c)\over c^{c+1}}\!\!\sum_{m=k+1}^\infty {c^m\over m!}
    \simeq {\Gamma(c+1,c)\over\Gamma(k+2)} c^{k-c},
\end{equation}
since the sum is strongly dominated in this limit by its first term.
Applying Stirling's approximation, $\Gamma(x) \simeq (x/\e)^x
\sqrt{2\pi/x}$, this gives
\begin{equation}
p_k \simeq {\Gamma(c+1,c)\over c^c}
           k^{-3/2} \e^k \biggl( {c\over k} \biggr)^k,
\end{equation}
which decays substantially faster asymptotically than any exponential.

\begin{figure}[t]
\includegraphics[width=\figurewidth]{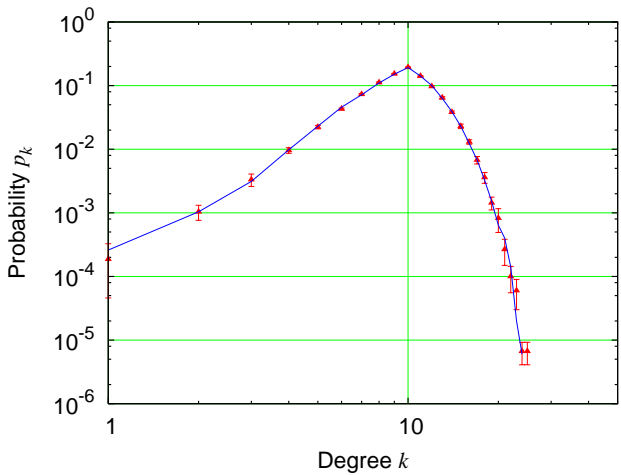}
\caption{The degree distribution of our model for the case of uniform
attachment ($\pi_k=\mbox{constant}$) with fixed size $n=50\,000$ and
$c=10$.  The points represent data from numerical simulations and the solid
line is the analytic solution.}
\label{fig:uniform}
\end{figure}

As a check on these calculations, we have performed extensive computer
simulations of the model.  In Fig.~\ref{fig:uniform} we show results for
the case $c=10$, along with the exact solution from Eqs.~\eqref{eq:pkkltc}
and~\eqref{eq:pkkgtc}.  As the figure shows, the agreement between the two
is excellent.

Before moving on to other issues, we note a different and particularly
simple case of a growing network with uniform attachment, the case in which
the vertices added have a Poisson degree distribution $c^k\e^{-c}/k!$ with
mean~$c$.  In that case the factor of $z^c$ in Eq.~\eqref{eq:uniconst} is
replaced with the generating function~$h(z)$ for the Poisson distribution:
\begin{equation}
h(z) = \sum_{k=0}^\infty {c^k\e^{-c}\over k!} z^k = \e^{c(z-1)},
\end{equation}
and the solution, Eq.~\eqref{eq:solngz1}, becomes
\begin{equation}
g(z) = {\e^{cz}\over1-z} \int_z^1 h(t)\,\e^{-ct} \>\d t = \e^{c(z-1)},
\end{equation}
which is itself the generating function for a Poisson distribution.  Thus
we see particularly clearly in this case that the equilibrium degree
distribution in the steady-state uniform attachment network is sharply
peaked with a Poisson tail.  In fact, the network in this case is simply an
uncorrelated random graph of the type famously studied by Erd\H{o}s and
R\'enyi~\cite{ER60}.  It is straightforward to see that if one starts with
such a graph and randomly adds and removes vertices with Poisson
distributed degrees, the graph remains an uncorrelated random graph with
the same degree distribution, and hence this distribution is necessarily
the fixed point of the evolution process, as the solution above
demonstrates.

\subsection{Preferential attachment and constant size}
\label{prefconst}
Our next example adds an extra degree of complexity to the picture: we
consider vertices that attach to others in proportion to their degree, the
so-called ``preferential attachment'' mechanism~\cite{BA99b}.  This implies
that our attachment kernel $\pi_k$ is linear in the degree, $\pi_k=Ak$ for
some constant~$A$.  The normalization requirement~\eqref{eq:normpik} then
implies that
\begin{equation}
\sum_{k=0}^\infty \pi_k p_k = A \sum_{k=0}^\infty k p_k = A\av{k} = 1,
\end{equation}
and hence $A=1/\av{k}$.  For the moment, let us continue to focus on the
case $r=1$ of constant network size, in which case $\av{k}=c$
(Eq.~\eqref{eq:avk}) and
\begin{equation}
\pi_k = {k\over c}.
\end{equation}
Then
\begin{equation}
f(z) = {1\over c} \sum_{k=0}^\infty kp_kz^k = {z\over c}\, g'(z),
\end{equation}
and Eq.~\eqref{eq:fg} becomes
\begin{equation}
{g(z)\over(1-z)^2} - {\d g\over\d z} = {z^c\over(1-z)^2} .
\end{equation}
The appropriate integrating factor in this case is $\e^{-1/(1-z)}$, which,
in conjunction with the boundary condition $g(1)=1$, gives
\begin{equation}
g(z) = e^{1/(1-z)} \int_z^1 \frac{t^c}{(1-t)^2} \e^{-1/(1-t)} \>\d t.
\end{equation}
Changing the variable of integration to $y=1/(1-t)$ this expression can be
written
\begin{eqnarray}
g(z) &=& \e^{1/(1-z)} \int_{1/(1-z)}^\infty
         \left( 1-\frac{1}{y} \right)^{\!c} \e^{-y} \,\d y \nonumber\\
     &=& \e^{1/(1-z)} \sum_{s=0}^c (-1)^s {c \choose s}
         \int_{1/(1-z)}^\infty \frac{\e^{-y}}{y^s} \,\d y \nonumber \\
     &=& 1 + \e^{1/(1-z)} \sum_{s=1}^c (-1)^s {c \choose s} \,
         \Gamma\!\left( 1-s, \frac{1}{1-z} \right). \qquad{}
\label{eq:prefg}
\end{eqnarray}
where $\Gamma(1-s,x) = \int_x^\infty e^{-y} \,y^{-s} \,\d y$ is again the
incomplete $\Gamma$-function, here appearing with a negative first
argument.

A useful identify for the case $s\ge1$ can be derived by integrating by
parts thus:
\begin{equation}
\Gamma(-s, x) = \frac{1}{s} \biggl[ \frac{\e^{-x}}{x^s} - \Gamma(1-s,x)
                \biggr].
\end{equation}
Iterating this expression then gives
\begin{equation}
\label{eq:gamident}
\Gamma(1-s, x) = - \frac{(-1)^s}{(s-1)!}
                 \biggl[ \Gamma(0,x) + \e^{-x}\!\sum_{m=1}^{s-1}
                 \frac{(-1)^m (m-1)!}{x^m} \biggr].
\end{equation}
($\Gamma(0,x) = \int_x^\infty (\e^{-y}/y) \,\d y$ is also known as the
exponential integral function $-\mathop{\mathrm{Ei}}(-x)$.)  Applying this
identity to~\eqref{eq:prefg} gives
\begin{eqnarray}
g(z) &=& 1 - \sum_{s=1}^c {c \choose s} \frac{1}{(s-1)!} \nonumber\\
     & & \hspace{-3em} {} \times
         \biggl[ \e^{1/(1-z)} \,\Gamma\!\biggl(0,\frac{1}{1-z}\biggr)
         + \sum_{m=1}^{s-1} (-1)^m (m-1)! \,(1-z)^m \biggr] \nonumber\\
     &=& q(z) - A_c \,\e^{1/(1-z)} \,\Gamma\!\biggl(0,\frac{1}{1-z}\biggr),
\end{eqnarray}
where $q(z)$ is a polynomial of degree $c-1$ and $A_c = \sum_{s=1}^c {c
\choose s}/(s-1)!$ depends only on $c$.  For $k \ge c$, then, the degree
distribution $p_k$ is given by the coefficients of $z^k$ in
$-A_c\,\e^{1/(1-z)}\,\Gamma\bigl(0,1/(1-z)\bigr)$.  We determine these
coefficients as follows.  Changing the variable of integration to
$x=y-z/(1-z)$, we find
\begin{equation}
-\e^{1/(1-z)} \,\Gamma\!\left(0,\frac{1}{1-z}\right) 
  = -\e \int_1^\infty \frac{\e^{-x}}{x+z/(1-z)} \,\d x.
\end{equation}
Then we expand the integrand to get
\begin{equation}
\frac{1}{x+z/(1-z)} = \frac{1}{x} - \sum_{k=1}^\infty
  \left( 1-\frac{1}{x} \right)^{\!k-1} \frac{z^k}{x^2}.
\end{equation}
Commuting the sum and the integral, we obtain
\begin{equation}
-\e^{1/(1-z)} \,\Gamma\!\left(0,\frac{1}{1-z}\right)
   = \sum_{k=0}^\infty a_k z^k,
\end{equation}
where
\begin{equation}
a_0 = -\e \int_1^\infty \frac{\e^{-x}}{x} \,\d x = -\e \,\Gamma(0,1),
\end{equation}
and for $k\ge1$
\begin{equation}
a_k = \e \int_1^\infty \left( 1-\frac{1}{x} \right)^{\!k-1}
      \frac{\e^{-x}}{x^2} \,\d x .
\label{eq:ak}
\end{equation}
Integrating by parts, we obtain a slightly simpler expression, 
\begin{equation} 
a_k = \frac{\e}{k} \int_1^\infty \left( 1-\frac{1}{x} \right)^{\!k}
      \e^{-x} \,\d x .
\label{eq:akint} 
\end{equation}

While the coefficients $a_k$ can be expressed exactly using hypergeometric
functions, a perhaps more informative approach is to employ a saddle-point
expansion.  The integrand of~\eqref{eq:akint} is unimodal in the interval
between 1 and $\infty$, and peaks at $x=\frac{1}{2}(1+\sqrt{4k+1})
\simeq \sqrt{k}$.  Approximating the integrand as a Gaussian around this
point, we obtain as $k \to \infty$
\begin{equation}
a_k \simeq \sqrt{\pi \e} \,k^{-3/4} \,\e^{-2 \sqrt{k}} 
\label{eq:akapprox}
\end{equation} 
and $p_k = A_c \,a_k$ for $k \ge c$ as stated above.

\begin{figure}[t]
\includegraphics[width=\figurewidth]{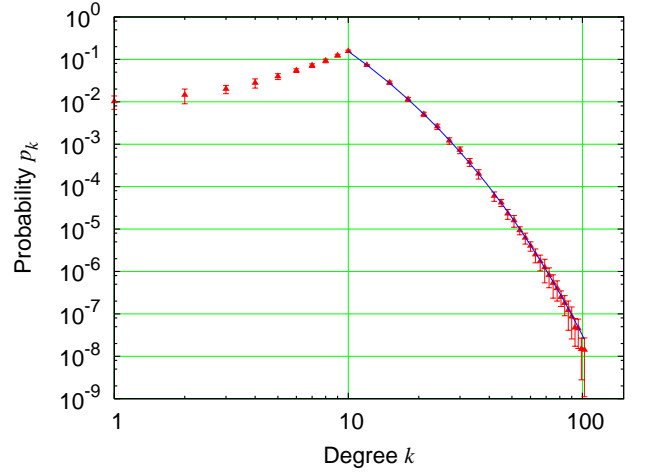}
\caption{The degree distribution for our model in the case of fixed size
$n=50\,000$ and $c=10$ with linear preferential attachment.  The points
represent data from our numerical simulations and the solid line is the
analytic solution for $k\ge c$.  Note that the tail of the distribution
does not follow a power law as in growing networks with preferential
attachment, but instead decays faster than a power law, as a stretched
exponential.}
\label{fig:pref}
\end{figure}

Figure~\ref{fig:pref} shows the form of this solution for the case $c=10$.
Also shown in the figure are results from computer simulations of the model
on systems of size $n=50\,000$ with $c=10$, which agree well with the
analytic results.  The appearance of the stretched exponential in
Eq.~\eqref{eq:akapprox} is worthy of note.  We are aware of only a few
cases of graphs with stretched exponential degree distributions that have
been discussed previously, for instance in growing networks with sublinear
preferential attachment~\cite{KR01} as well as in empirical network
data~\cite{NFB02}.

\subsection{Preferential attachment in a growing network}
\label{sec:prefgrow}
We now come to the third and most complex of our example networks, in which
we combine preferential attachment with net growth of the network, $r<1$.
(Logically, we should perhaps first solve the case of a growing network
without preferential attachment, which in fact we have done.  But the
solution turns out to have no qualitatively new features to distinguish it
from the constant size case and is mathematically tedious besides.  Given
the large amount of effort it requires and its modest rewards, therefore,
we prefer to skip this case and move on to more fertile ground.)

As before, perfect linear preferential attachment implies $\pi_k=k/\av{k}$
or
\begin{equation}
\pi_k = \half(1+r) {k\over c},
\end{equation}
where we have made use of Eq.~\eqref{eq:avk}.  Then $f(z) = (1+r)zg'(z)/2c$
and Eq.~\eqref{eq:fg} becomes
\begin{equation}
g(z) - (1-z)\bigl[r-\half(1+r)z\bigr]\,{\d g\over\d z} = z^c.
\label{eq:lastreq}
\end{equation}
An integrating factor for the left-hand side in this case is
$\bigl|(\alpha-z)/(1-z)\bigr|^{-2/(1-r)}$ where $\alpha=2r/(1+r)$.  (Note
that $\alpha<1$ when $r<1$.)  Unfortunately, this integrating factor is
non-analytic at $z=\alpha$, which makes integrals traversing this point
cumbersome.  To circumvent this difficulty, we observe that the second term
in Eq.~\eqref{eq:lastreq} vanishes at $z=\alpha$, giving
$g(\alpha)=\alpha^c$.  This provides us with an alternative boundary
condition on~$g(z)$, allowing us to fix the integrating constant while only
integrating up to~$z=\alpha$.  It is then straightforward to show that
\begin{eqnarray}
g(z) &=& {2\over1+r} \biggl( {\alpha-z\over1-z} \biggr)^{-2/(1-r)}\nonumber\\
     & & \quad{}\times
       \int_z^\alpha \biggl( {\alpha-t\over1-t} \biggr)^{2/(1-r)} \!
       {t^c \>\d t\over(1-t)(\alpha-t)},\quad
\end{eqnarray}
for $z\le\alpha$.  Since the degree distribution is entirely determined by
the behavior of $g(z)$ at the origin, it is adequate to restrict our
solution to this regime.

Changing variables to $u = (\alpha-t)/(1-\alpha)$, we find
\begin{eqnarray}
g(z) &=& {2\over1+r} \biggl( {\alpha-z\over1-z} \biggr)^{1-\gamma}
         (1-\alpha)^{-1} \nonumber\\
     & & \times \int_0^{{\alpha-z\over1-\alpha}}
         \biggl( {u\over1+u} \biggr)^\gamma
         \bigl[ \alpha - (1-\alpha)u \bigr]^c \>{\d u\over u^2},
\end{eqnarray}
where $\gamma = (3-r)/(1-r)$.  If we expand the last factor in the
integrand, this becomes
\begin{eqnarray}
g(z) &=& {2\over1+r} \sum_{s=0}^c (-1)^s {c\choose s}
         (1-\alpha)^{s-1} \alpha^{c-s} \nonumber\\
     & & \quad{}\times \biggl( {\alpha-z\over1-z} \biggr)^{1-\gamma}
         \int_0^{{\alpha-z\over1-\alpha}}
         {u^{s+\gamma-2}\over(1+u)^\gamma} \>\d u.\qquad
\label{eq:finalgz}
\end{eqnarray}
\begin{widetext}
We observe the following useful identity:
\begin{equation}
\int_0^x {u^\beta\over(1+u)^\gamma} \>\d u
  = \int_0^x \biggl( {u\over1+u} \biggr)^\beta (1+u)^{\beta-\gamma} \>\d u
  = {x^\beta\over(\beta-\gamma+1)(1+x)^{\gamma-1}}
    - {\beta\over\beta-\gamma+1} \int_0^x
    {u^{\beta-1}\over(1+u)^\gamma} \>\d u,
\label{eq:useful}
\end{equation}
where the second equality is derived via integration by parts.  Setting
$\beta=s+\gamma-2$ and $x=(\alpha-z)/(1-\alpha)$ and noting that the last
integral has the same form as the first, we can employ this identity
iteratively $s-1$ times to get
\begin{eqnarray}
\biggl( {\alpha-z\over1-z} \biggr)^{1-\gamma}
    \int_0^{{\alpha-z\over1-\alpha}}
    {u^{s+\gamma-2}\over(1+u)^\gamma} \>\d u
  &=& (-1)^{s+1} {\Gamma(s+\gamma-1)\over\Gamma(s)} \nonumber\\
  & & \quad{}\times \biggl[ {1\over\Gamma(\gamma)}
    \biggl( {\alpha-z\over1-z} \biggr)^{1-\gamma}
    \!\int_0^{{\alpha-z\over1-\alpha}} {u^{\gamma-1}\over(1+u)^\gamma} \>\d u
    + \sum_{m=1}^{s-1} {(-1)^m\over\Gamma(\gamma+m)}
      \biggl( {\alpha-z\over1-z} \biggr)^m \biggr].\qquad
\end{eqnarray}
\end{widetext}
The final sum can be evaluated in closed form in terms of the incomplete
$\Gamma$-function, but our primary focus here is on the preceding term.
Substituting into Eq.~\eqref{eq:finalgz}, we see that $g(z) = q(z) +
A_{c,r} h(z)$, where
\begin{equation}
\label{eq:hz}
h(z)    = - \biggl( \frac{\alpha-z}{1-z} \biggr)^{1-\gamma}
            \int_0^{{\alpha-z\over1-\alpha}} {u^{\gamma-1}\over(1+u)^\gamma}
            \>\d u,
\end{equation}
\begin{equation}
\label{eq:acr}
A_{c,r} = {2\over1+r} \sum_{s=0}^c {c\choose s}
            (1-\alpha)^{s-1} \alpha^{c-s}
            \frac{\Gamma(\gamma+s-1)}{\Gamma(\gamma)\Gamma(s)},
\end{equation}
and $q(z)$ is a polynomial of order $c-1$ in~$z$.

Since $A_{c,r}$ depends only on $c$ and~$r$ and $q(z)$ has no terms in $z$
of order $z^c$ or higher, the degree distribution for $k\ge c$ is, to
within a multiplicative constant, given by the coefficients in the
expansion of $h(z)$ about zero.  Making the change of variables
\begin{equation}
u = {y\over(1-z)/(\alpha-z)-y},
\end{equation}
we find that
\begin{equation}
h(z) = - \int_0^1 {y^{\gamma-1}\>\d y\over(1-z)/(\alpha-z)-y},
\end{equation}
and expanding the integrand in powers of~$z$ we obtain $h(z) =
\sum_{k=0}^\infty a_k z^k$ with
\begin{eqnarray}
a_k &=& (1-\alpha) \int_0^1
{(1-y)^{k-1}\over(1-\alpha y)^{k+1}}\,y^{\gamma-1} \>\d y \nonumber\\
    &=& {\gamma-1\over k} \int_0^1
      \biggl( {1-y\over 1-\alpha y} \biggr)^k y^{\gamma-2} \>\d y,
\label{eq:ak2}
\end{eqnarray}
for $k\ge1$, where the second equality follows via an integration by parts.

As in the case of constant size, we can express these coefficients in
closed form using special functions, but if we are primarily interested in
the form of the tail of the degree distribution then a more revealing
approach is to make a further substitution $y=x/k$, giving
\begin{equation}
a_k = (\gamma-1)\,k^{-\gamma} \int_0^k {(1-x/k)^k\over(1-\alpha x/k)^k}\,
      x^{\gamma-2} \>\d x.
\end{equation}
In the limit of large~$k$ this becomes
\begin{eqnarray}
a_k &\simeq& (\gamma-1)\,k^{-\gamma} \int_0^\infty \e^{-(1-\alpha)x}
           x^{\gamma-2} \>\d x \nonumber\\
    &=& { \Gamma(\gamma)\over(1-\alpha)^{\gamma-1}}\,k^{-\gamma},
\label{eq:akapprox2}
\end{eqnarray}
and $p_k = A_{c,r} \,a_k$ for $k \ge c$ as stated above.

Thus the tail of the degree distribution follows a power law with
exponent~$\gamma=(3-r)/(1-r)$.  Note that this exponent diverges as $r\to1$
so that the power law becomes ever steeper as the growth rate slows,
eventually assuming the stretched exponential form of
Eq.~\eqref{eq:akapprox}---steeper than any power law---in the limit $r=1$.
In the limit $r\to0$ we recover the established power-law behavior $a_k\sim
k^{-3}$ for growing graphs with preferential attachment and no vertex
removal~\cite{Price76,BA99b,DMS00,KRL00}.

\begin{figure}[t]
\includegraphics[width=\figurewidth]{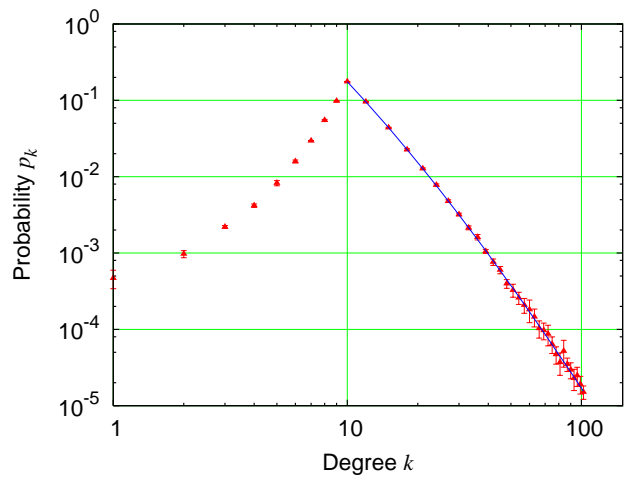}
\caption{Degree distribution for a growing network with linear
preferential attachment and $r=\half$, $c=10$.  The solid line represents
the analytic solution, Eqs.~\eqref{eq:acr}, and~\eqref{eq:ak2}, for $k\ge
c$ and the points represent simulation results for systems with final size
$n=100\,000$ vertices.}
\label{fig:prefgrow}
\end{figure}

In Fig.~\ref{fig:prefgrow} we show the form of the degree distribution for
this model for the case $r=\half$, $c=10$, along with numerical results
from simulations of the model on networks of (final) size $n=100\,000$
vertices.  The power-law behavior is clearly visible on the logarithmic
scales used as a straight line in the tail of the distribution.  Once again
the analytic solution and simulations are in excellent agreement.

We note that Chung and Lu~\cite{CL04} and Cooper, Frieze, and
Vera~\cite{CFV04} have independently demonstrated power-law behavior in the
degree distribution of growing networks, using models similar to ours.
Their results are asymptotic approximations describing the tail of the
distribution, rather than exact solutions, but they find the same
dependence of the exponent on the growth rate.

\section{Discussion}
In this paper we have studied models of the time evolution of networks in
which, in addition to the widely considered case of addition of vertices,
we also include vertex removal.  We have given exact solutions for cases in
which vertices are added and removed at the same rate, a potential model
for steady-state networks such as peer-to-peer networks, and cases in which
the rate of addition exceeds the rate of removal, which we regard as a
simple model for the growth of, for example, the worldwide web.

We find very different behaviors in these various cases.  For a
steady-state network in which newly added vertices attach to others at
random we find a degree distribution, Eqs.~\eqref{eq:pkkltc}
and~\eqref{eq:pkkgtc}, which is sharply peaked about its maximum and has a
rapidly decaying (Poisson) tail.  This distribution is quite unlike the
right-skewed degree distributions found in many real-world networks, but as
a possible form for a ``designed'' network such as a peer-to-peer network
it might be preferable over skewed forms, being more homogeneous and hence
distributing traffic more evenly.

If newly appearing vertices attach to others using a linear preferential
attachment mechanism, whereby vertices gain new edges in proportion to the
number they already possess, we find that the degree distribution becomes a
stretched exponential, Eqs.~\eqref{eq:akint} and~\eqref{eq:akapprox}, a
substantially broader distribution than that of the random attachment case,
though still more rapidly decaying than the power laws often seen in
growing networks.

And in the case where the network shows net growth, adding vertices faster
than it loses them, we find that the degree distribution follows a power
law, Eqs.~\eqref{eq:ak2} and~\eqref{eq:akapprox2}, with an exponent
$\gamma$ that assumes values in the range $3\le\gamma<\infty$, diverging as
the growth rate tends to zero.

This last result is of interest for a number of reasons.  First, it shows
that power-law behavior can be rigorously established in networks that grow
but also lose vertices.  Most previous analytic models of network growth
have focused solely on vertex addition.  And while the real worldwide web
and other networks appear to have degree distributions that closely follow
power laws, these networks also clearly lose vertices as well as gaining
them.  The results presented here demonstrate that the widely studied
mechanism of preferential attachment for generating power-law behavior also
works in this regime.

On the other hand, the large values of the exponent~$\gamma$ generated by
our model appear not to be in agreement with the behavior observed in
real-world networks, most of which have exponents in the range from 2
to~3~\cite{AB02,DM02,Newman03d}.  There are well-known mechanisms that can
reduce the exponent from 3 to values slightly lower---specifically the
generalization of the preferential attachment model to the case of a
directed network~\cite{Price76,DMS00}, which is in any case a more
appropriate model for the worldwide web.  In the limit of low growth rate,
however, our model predicts a diverging exponent and, while the exact value
may not be accurate because of a host of complicating factors, it seems
likely that the divergence itself is a robust phenomenon; as other authors
have commented, there are good reasons to believe that net growth is one of
the fundamental requirements for the generation of power-law degree
distributions by the kind of mechanisms considered here.

Thus the fact that we do not observe very large exponents in real networks
appears to indicate that most networks are in a regime where growth
dominates over vertex loss by a wide margin.  It is possible however that
this will not always be the case.  The web, for example, has certainly
being enjoying a period of very vigorous growth since its appearance in the
early 1990s, but it could be that this is a sign primarily of its youth,
and that as the network matures its size will grow more slowly, the
vertices added being more nearly balanced by those taken away.  Were this
to happen, we would expect to see the exponent of the degree distribution
grow larger.  A sufficiently large exponent would make the distribution
indistinguishable experimentally from an exponential or stretched
exponential distribution, although we do not realistically anticipate
seeing behavior of this type any time in the near future.

\begin{acknowledgments}
This work was funded in part by the National Science Foundation under
grants DMS--0234188 and PHY--0200909 and by the James S. McDonnell
Foundation.  CM thanks the University of Michigan and the Center for the
Study of Complex Systems for their hospitality, and Tracy Conrad and
Rosemary Moore for their support.
\end{acknowledgments}

\end{document}